\documentclass[11pt,a4paper]{article}

\usepackage{amssymb}
\usepackage{amsmath}
\usepackage{epsfig}
\usepackage{graphicx}
\usepackage{enumerate}
\usepackage[footnotesize]{caption}
\newcommand{\mathsym}[1]{{}}

\newcommand{\ket}[1]{\left|#1\right\rangle}      
\newcommand{\bra}[1]{\left\langle #1\right|}     

\begin{document}

\newpage
\setcounter{page}{0}
\begin{titlepage}
\begin{flushright}
{\footnotesize AEI - 2010 - 154}
\end{flushright}
\vskip 1.5cm
\begin{center}
{\Large \textbf{A new representation for the partition function of the  \\ six vertex model with domain wall boundaries}}\\
\vspace{1.5cm}
{\large W. Galleas} \\
\vspace{2cm}
{\it Max-Planck-Institut f\"ur Gravitationsphysik \\
Albert-Einstein-Institut \\
Am M\"uhlenberg 1, 14476 Potsdam, Germany}\\
\end{center}
\vspace{1.5cm}

\begin{abstract}
We obtain a new representation for the partition function of the six vertex model with domain wall boundaries
using a functional equation recently derived by the author. This new
representation is given in terms of a sum over the permutation group where the partial homogeneous limit can be taken
trivially. We also show by construction that this partition function satisfies a linear partial differential
equation. 
\end{abstract}

\vspace{2.0cm}
\centerline{{\small PACS numbers:  05.50+q, 02.30.IK}}
\vspace{.1cm}
\centerline{{\small Keywords: Yang-Baxter algebra, Domain wall boundaries, Functional equations,}}
\centerline{{\small Partial differential equations}}
\vspace{1.5cm}
\centerline{{\small October 2010}}
\end{titlepage}

\tableofcontents

\section{Introduction}
In the past decades the mathematical structure underlying integrable systems has been intensively studied and
even after many years it seems that its richness has not yet been completely exhausted. 
Integrable systems can be solved by a variety of methods, ranging from functional to algebraic
methods, whose solution is usually expressed in terms of solutions of Bethe ansatz equations \cite{BET,LIE,BAX,qism2}. 
By way of contrast, the six vertex model with domain wall boundary conditions (DWBC) is one 
exception for this general behaviour, and in fact the computation of its partition function and free-energy
does not rely in Bethe ansatz like solutions \cite{KOR,IZE,JUS}. 

In the pioneer work \cite{KOR} it was demonstrated that this model obeys a 
recurrence relation relating the partition function of the system on a square lattice
of size $L \times L$ to the one on a lattice of size $(L-1) \times (L-1)$. Five years later
Izergin proposed a determinant solution for this recurrence relation which, together with
extra properties, determines uniquely the partition function of the system \cite{IZE}.

On the other hand it is nowadays well understood the fundamental role of the Yang-Baxter equation
and the Yang-Baxter algebra in the construction and solution of integrable systems, 
though the Yang-Baxter algebra does not seem to play any explicit role in Izergin-Korepin
solution of the six vertex model with DWBC. Recently it was proposed an alternative
approach for computing this partition function where the Yang-Baxter algebra is the main 
ingredient \cite{GAL}. The approach of \cite{GAL} makes explicit use of the Yang-Baxter algebra in order
to derive a functional equation determining the partition function of the six vertex model with DWBC.
Though without a rigorous proof, in the framework of \cite{GAL} the partition function
is uniquely determined by three conditions:
\begin{enumerate}[(i)]
\item Functional equation
\item Polynomial structure
\item Asymptotic behaviour
\end{enumerate}
and here we aim to demonstrate that Korepin's recurrence relation can be suitably introduced
in this framework removing the need of polynomial solutions. This is of particular interest
for further applications since there exist relevant models, such as the Hubbard model, whose $R$-matrix
contains non-polynomial elements \cite{SHA}. Moreover, the introduction of the recurrence relation
in the functional equation derived in \cite{GAL} yields naturally an explicit representation for the
partition function of the six vertex model with DWBC whose homogeneous limit for the vertical degrees
of freedom can be obtained trivially.  
In a second analysis we also complement the results of \cite{GAL} by showing that the functional equation previously 
obtained for the partition function of the six vertex model with DWBC can be converted into a linear partial differential equation.

This paper is organized as follows. In the section 2 we recall some basic definitions and results of \cite{GAL} as a starting point. 
In the section 3 we demonstrate how Korepin's recurrence relation can fit in our functional equation yielding a
novel representation for the partition function of the six vertex model with DWBC. 
In the section 4 we recast the functional
equation for the partition function as an operator equation followed by its differential representation. 
Final comments and concluding remarks are discussed in the section 5.

\section{Functional relations and domain wall boundaries}
In this section we recall some previous results and definitions associated with the derivation of a functional
equation determining the partition function of the six vertex model with domain wall boundaries. Let us consider 
the matrix $\mathcal{L} (\lambda)$ given by
\begin{equation}
\label{lm}
\mathcal{L} (\lambda)= \left( \begin{matrix}
a(\lambda) & 0 & 0 & 0 \\
0 & b(\lambda) & c(\lambda) & 0 \\
0 & c(\lambda) & b(\lambda) & 0 \\
0 & 0 & 0 & a(\lambda)
\end{matrix} \right) 
\end{equation}
containing the statistical weights of the six vertex model. The non-null entries are given by
$a(\lambda)=\sinh{(\lambda + \gamma)}$, $b(\lambda)=\sinh{(\lambda)}$ and $c(\lambda)=\sinh{(\gamma)}$, where $\lambda$ 
and $\gamma$ are complex variables. The matrix $\mathcal{L} (\lambda)$ satisfies the Yang-Baxter equation, namely
\begin{equation}
\label{yb}
\mathcal{L}_{12}(\lambda - \mu) \mathcal{L}_{13}(\lambda - \nu) \mathcal{L}_{23}(\mu - \nu) =   
\mathcal{L}_{23}(\mu - \nu) \mathcal{L}_{13}(\lambda - \nu) \mathcal{L}_{12}(\lambda - \mu) 
\end{equation}
where $\mathcal{L}_{ij} \in \mbox{End} \left( V_i \otimes V_j \right)$ and $V_i \cong \mathbb{C}^2$, and
in this way the monodromy matrix $\mathcal{T}$ defined by
\begin{equation}
\label{mono}
\mathcal{T}(\lambda, \{ \mu_{k} \}) = \mathcal{L}_{\mathcal{A} 1}(\lambda - \mu_{1}) \mathcal{L}_{\mathcal{A} 2}(\lambda - \mu_{2}) \dots \mathcal{L}_{\mathcal{A} L}(\lambda - \mu_{L}) 
\end{equation}
satisfies the following quadratic relation usually referred as Yang-Baxter algebra
\begin{equation}
\label{rtt}
R(\lambda - \nu) \; \mathcal{T}(\lambda, \{ \mu_{k} \}) \otimes \mathcal{T}(\nu, \{ \mu_{k} \}) = 
\mathcal{T}(\nu, \{ \mu_{k} \}) \otimes \mathcal{T}(\lambda, \{ \mu_{k} \}) \; R(\lambda - \nu) \; .
\end{equation}
Here $R(\lambda) = P \mathcal{L}(\lambda)$ and $P$ denotes the standard permutation matrix. 

The monodromy matrix $\mathcal{T}(\lambda , \{ \mu_{k} \} )$ can be conveniently written in terms
of operators $A(\lambda , \{ \mu_{k} \} )$, $B(\lambda , \{ \mu_{k} \} )$, $C(\lambda , \{ \mu_{k} \} )$
and $D(\lambda , \{ \mu_{k} \} )$, i.e.
\begin{equation}
\label{abcd}
\mathcal{T}(\lambda , \{ \mu_{k} \} ) = \left( \begin{matrix}
A(\lambda , \{ \mu_{k} \} ) & B(\lambda , \{ \mu_{k} \} ) \\
C(\lambda , \{ \mu_{k} \} ) & D(\lambda , \{ \mu_{k} \} )
\end{matrix} \right) \; ,
\end{equation}
and as demonstrated in \cite{KOR} the partition function $Z$ of the six vertex model with domain wall boundaries
in a lattice of size $L \times L$  can be expressed as
\begin{equation}
\label{pf}
Z = \bra{\bar{0}} \prod_{j=1}^L B(\lambda_{j}, \{ \mu_k \} ) \ket{0}
\end{equation}
where the states $\ket{0}$ and $\ket{\bar{0}}$ consist of the ferromagnetic states
\begin{eqnarray}
\label{vac}
\ket{0} = \bigotimes_{i=1}^{L} 
\left( \begin{matrix}
1 \\
0 
\end{matrix} \right)
\;\;\; \mbox{and} \;\;\;
\ket{\bar{0}} = \bigotimes_{i=1}^{L} 
\left( \begin{matrix}
0 \\
1 
\end{matrix} \right).
\end{eqnarray}

Among the three conditions given in \cite{GAL} determining the partition function $Z$, the main one consist of the functional equation
\begin{eqnarray}
\label{FZ}
&&\sum_{i=1}^{L+1} M_i \; Z(\lambda_1, \dots, \lambda_{i-1}, \lambda_{i+1}, \dots, \lambda_{L+1} ) \nonumber \\
&+& \sum_{1 \leq i < j \leq L+1} N_{ji} \; Z(\lambda_0 , \lambda_1, \dots, \lambda_{i-1}, \lambda_{i+1}, \dots, \lambda_{j-1}, \lambda_{j+1}, \dots, \lambda_{L+1} ) =0 
\end{eqnarray}
where we write $Z(\lambda_1, \dots, \lambda_{L} )$ for the partition function (\ref{pf}) omitting the dependence with the variables $\{ \mu_j \}$. In their turn the coefficients
$M_i$ and $N_{ji}$ are given by
\begin{eqnarray}
\label{mn}
M_i &=& \frac{c(\lambda_i - \lambda_0)}{b(\lambda_i - \lambda_0)} \prod_{l=1}^{L} a(\lambda_0 - \mu_l) b(\lambda_i - \mu_l) \prod_{\stackrel{k=1}{k \neq i}}^{L+1} \frac{a(\lambda_i - \lambda_k)}{b(\lambda_i - \lambda_k)} \frac{a(\lambda_k - \lambda_0)}{b(\lambda_k - \lambda_0)} \nonumber \\
&+& \frac{c(\lambda_0 - \lambda_i)}{b(\lambda_0 - \lambda_i)} \prod_{l=1}^{L} a(\lambda_i - \mu_l) b(\lambda_0 - \mu_l) \prod_{\stackrel{k=1}{k \neq i}}^{L+1} \frac{a(\lambda_0 - \lambda_k)}{b(\lambda_0 - \lambda_k)} \frac{a(\lambda_k - \lambda_i)}{b(\lambda_k - \lambda_i)}
\end{eqnarray} 
\begin{eqnarray}
\label{mn1}
N_{ji} &=& \frac{c(\lambda_0 - \lambda_j)}{b(\lambda_0 - \lambda_j)} \frac{c(\lambda_i - \lambda_0)}{b(\lambda_i - \lambda_0)} \frac{a(\lambda_j - \lambda_i)}{b(\lambda_j - \lambda_i)} \prod_{l=1}^{L} a(\lambda_i - \mu_l) b(\lambda_j - \mu_l) \prod_{\stackrel{m=1}{m \neq i,j}}^{L+1} \frac{a(\lambda_j - \lambda_m)}{b(\lambda_j - \lambda_m)} \frac{a(\lambda_m - \lambda_i)}{b(\lambda_m - \lambda_i)} \nonumber \\
&+& \frac{c(\lambda_0 - \lambda_i)}{b(\lambda_0 - \lambda_i)} \frac{c(\lambda_j - \lambda_0)}{b(\lambda_j - \lambda_0)} \frac{a(\lambda_i - \lambda_j)}{b(\lambda_i - \lambda_j)} \prod_{l=1}^{L} a(\lambda_j - \mu_l) b(\lambda_i - \mu_l) \prod_{\stackrel{m=1}{m \neq i,j}}^{L+1} \frac{a(\lambda_i - \lambda_m)}{b(\lambda_i - \lambda_m)} \frac{a(\lambda_m - \lambda_j)}{b(\lambda_m - \lambda_j)}. \nonumber \\
\end{eqnarray}

Here we shall also employ the variables $x_i = e^{2(\lambda_i - \mu_i)}$ in order to characterize the polynomial 
structure of $Z$. More precisely, the partition function (\ref{pf}) exhibits the following polynomial structure
\begin{equation}
\label{pol}
\displaystyle Z(\lambda_1 , \dots , \lambda_L) = \frac{\bar{Z}(x_1 , \dots , x_L)}{\displaystyle \prod_{i=1}^{L} x_{i}^{\frac{L-1}{2}}}
\end{equation}
where $\bar{Z}(x_1 , \dots , x_L)$ is a polynomial of degree $L-1$ in each variable $x_i$ separately.

Besides the two conditions discussed above, the full determination of the partition function also makes use of the asymptotic behaviour
$\bar{Z}(x_1 , \dots , x_L) \sim \frac{(q-q^{-1})^L}{2^{L^2}} [ L ]_{q^2} ! \; (x_{1} \dots x_{L})^{L-1} $ as $x_i \rightarrow \infty$, where
$[ L ]_{q^2} !$ denotes the $q$-factorial function defined as
\begin{equation}
[ L ]_{q^2} ! = 1 (1+q^2)(1+q^2+q^4) \dots (1 + q^2 + \dots + q^{2(L-1)}) \; .
\end{equation}

The partition function (\ref{pf}) also exhibits some extra properties besides the three conditions discussed above. For instance,
$Z$ is a symmetric function under the exchange of variables $\lambda_i \leftrightarrow \lambda_j$, i.e.
\begin{equation}
\label{sym}
Z(\lambda_1 , \dots , \lambda_i , \dots , \lambda_j , \dots , \lambda_L ) = Z(\lambda_1 , \dots , \lambda_j , \dots , \lambda_i , \dots , \lambda_L ), 
\end{equation}
which follow directly from the commutation relation
\begin{equation}
B(\lambda , \{ \mu_k \})  B(\nu , \{ \mu_k \}) =  B(\nu , \{ \mu_k \}) B(\lambda , \{ \mu_k \})
\end{equation}
encoded in the relation (\ref{rtt}). 

We close this section remarking that the partition function (\ref{pf}) is also symmetric under the exchange of variables
$\mu_i \leftrightarrow \mu_j$ as discussed in \cite{KOR}, however this property will not be required for our forthcoming analysis.

\section{A new representation for $Z(\lambda_1, \dots, \lambda_{L} )$}
In this section we aim to demonstrate how we can combine the functional equation (\ref{FZ}) and 
Korepin's recurrence relation \cite{KOR} in order to produce a new representation for the partition function
$Z$. As a matter of fact it turns out that the functional equation and the recurrence relation
together are enough to completely determine the partition function.

In order to construct a representation for the partition function $Z$ we firstly recall
the recurrence relation derived in \cite{KOR} adjusted to the notation
we are considering here. 

Since this recurrence relation has been already discussed in many works \cite{qism2,KOR,KOR1} we shall
omit its derivation and here we only need the relation
\begin{eqnarray}
\label{RR}
\left. Z(\lambda_1, \dots, \lambda_{L} | \mu_1, \dots, \mu_{L} ) \right|_{\lambda_1 = \mu_1} = c \prod_{j=2}^{L} a(\lambda_j - \mu_1) a(\mu_1 - \mu_j)
Z(\lambda_2, \dots, \lambda_{L} | \mu_2, \dots, \mu_{L} ) \nonumber \\
\end{eqnarray} 
with $Z(\lambda_1 | \mu_1 ) = c(\lambda_1 - \mu_1 )$. In the relation (\ref{RR}) we have included explicitly the dependence 
of $Z$ with the variables $\{ \mu_j \}$ and
also have assumed that $Z(\lambda_1, \dots, \lambda_{n} | \mu_1, \dots, \mu_{n} )$ denotes the partition function $Z$ on a lattice of size $n \times n$.
In this way for instance, the Eq. (\ref{RR}) is a first order recurrence relation connecting the partition function on a $(L-1)\times (L-1)$ lattice to the one
on a $L \times L$ lattice for a particular value of the variable $\lambda_1$.

For ilustrative purposes we shall first consider the functional equation (\ref{FZ}) with $L=2$
which is then given by
\begin{eqnarray}
\label{L2}
&&M_1 Z(\lambda_2 , \lambda_3 | \mu_1 , \mu_2) + M_2 Z(\lambda_1 , \lambda_3 | \mu_1 , \mu_2) + M_3 Z(\lambda_1 , \lambda_2 | \mu_1 , \mu_2) \nonumber \\
&&+ N_{21} Z(\lambda_0 , \lambda_3 | \mu_1 , \mu_2) + N_{31} Z(\lambda_0 , \lambda_2 | \mu_1 , \mu_2) + N_{32} Z(\lambda_0 , \lambda_1 | \mu_1 , \mu_2)
= 0
\end{eqnarray}
where the coefficients $M_i$ and $N_{ji}$ follows from (\ref{mn}) and (\ref{mn1}) with the appropriate value of $L$. 

By setting $\lambda_0 = \mu_1$ and $\lambda_3 = \mu_1 - \gamma$ we obtain
\begin{equation}
\left. M_1 \right|_{\lambda_0 = \mu_1, \lambda_3 = \mu_1 - \gamma} = \left. M_2 \right|_{\lambda_0 = \mu_1, \lambda_3 = \mu_1 - \gamma} = 0
\end{equation}
while $M_3$, $N_{21}$, $N_{31}$ and $N_{32}$ remain finite. More precisely we have
\begin{eqnarray}
\left. M_3 \right|_{\lambda_0 = \mu_1, \lambda_3 = \mu_1 - \gamma} &=& - c^2 a(\mu_1 - \mu_2) a(\mu_2 - \mu_1) \nonumber \\  
\left. N_{31} \right|_{\lambda_0 = \mu_1, \lambda_3 = \mu_1 - \gamma} &=& c^2 a(\mu_2 - \mu_1) a(\lambda_1 - \mu_2) \frac{b(\lambda_2 - \mu_1)}{a(\lambda_2 - \mu_1)}
\frac{a(\lambda_2 - \lambda_1)}{b(\lambda_2 - \lambda_1)} \nonumber \\
\left. N_{32} \right|_{\lambda_0 = \mu_1, \lambda_3 = \mu_1 - \gamma} &=& c^2 a(\mu_2 - \mu_1) a(\lambda_2 - \mu_2) \frac{b(\lambda_1 - \mu_1)}{a(\lambda_1 - \mu_1)}
\frac{a(\lambda_1 - \lambda_2)}{b(\lambda_1 - \lambda_2)} 
\end{eqnarray}
and we omit the explicit form of $\left. N_{21} \right|_{\lambda_0 = \mu_1, \lambda_3 = \mu_1 - \gamma}$ since it will not be required.

For this particular choice of variables $\lambda_0$ and $\lambda_3$ the Eq. (\ref{L2}) is then reduced to 
\begin{eqnarray}
\label{L2a}
Z(\lambda_1 , \lambda_2 | \mu_1 , \mu_2) &=&   \frac{a(\lambda_1 - \mu_2)}{a(\mu_1 - \mu_2)} \frac{b(\lambda_2 - \mu_1)}{a(\lambda_2 - \mu_1)}
\frac{a(\lambda_2 - \lambda_1)}{b(\lambda_2 - \lambda_1)}  Z(\mu_1 , \lambda_2 | \mu_1 , \mu_2) \nonumber \\
&+& \frac{a(\lambda_2 - \mu_2)}{a(\mu_1 - \mu_2)} \frac{b(\lambda_1 - \mu_1)}{a(\lambda_1 - \mu_1)}
\frac{a(\lambda_1 - \lambda_2)}{b(\lambda_1 - \lambda_2)} Z(\mu_1 , \lambda_1 | \mu_1 , \mu_2) \nonumber \\
&-& Z(\mu_1 , \mu_1 - \gamma | \mu_1 , \mu_2) \cdot \left. \frac{N_{21}}{M_3} \right|_{\lambda_0 = \mu_1, \lambda_3 = \mu_1 - \gamma} \; ,
\end{eqnarray} 
and if we set $\lambda_2 = \mu_1$ in (\ref{L2a}) and consider that the partition function $Z$ is symmetric under the exchange of variables
$\lambda_i \leftrightarrow \lambda_j$, we are left with the following identity
\begin{equation}
\label{zero}
\left\{ \left. \frac{N_{21}}{M_3} \right|_{\stackrel{\lambda_0 = \mu_1, \lambda_3 = \mu_1 - \gamma}{\lambda_2 = \mu_1}} \right\}  \cdot  Z(\mu_1 , \mu_1 - \gamma | \mu_1 , \mu_2) = 0 \; .
\end{equation}
Since the quantity inside the brackets in (\ref{zero}) is finite, we can conclude that
$Z(\mu_1 , \mu_1 - \gamma | \mu_1 , \mu_2) = 0$. Thus the relation (\ref{L2a}) simplifies to
\begin{eqnarray}
\label{L2b}
Z(\lambda_1 , \lambda_2 | \mu_1 , \mu_2) &=&   \frac{a(\lambda_1 - \mu_2)}{a(\mu_1 - \mu_2)} \frac{b(\lambda_2 - \mu_1)}{a(\lambda_2 - \mu_1)}
\frac{a(\lambda_2 - \lambda_1)}{b(\lambda_2 - \lambda_1)}  Z(\mu_1 , \lambda_2 | \mu_1 , \mu_2) \nonumber \\
&+& \frac{a(\lambda_2 - \mu_2)}{a(\mu_1 - \mu_2)} \frac{b(\lambda_1 - \mu_1)}{a(\lambda_1 - \mu_1)}
\frac{a(\lambda_1 - \lambda_2)}{b(\lambda_1 - \lambda_2)} Z(\mu_1 , \lambda_1 | \mu_1 , \mu_2) \; .
\end{eqnarray}
Notice however that the property $Z(\mu_1 , \mu_1 - \gamma | \mu_1 , \mu_2) = 0$ could also have been inferred from the recurrence relation (\ref{RR}). 

Now we can simply insert the recurrence relation (\ref{RR}) into the relation (\ref{L2a}), and by doing so we automatically obtain 
\begin{equation}
\label{z2}
Z(\lambda_1 , \lambda_2 | \mu_1 , \mu_2) = F_{12} + F_{21}
\end{equation}
where
\begin{equation}
\label{f2}
F_{ij} = c^2 a(\lambda_i - \mu_2) b(\lambda_j - \mu_1) \frac{a(\lambda_j - \lambda_i)}{b(\lambda_j - \lambda_i)} \; .
\end{equation}

This procedure can be straightforwardly extended to the case $L=3$. In that case 
the functional equation (\ref{FZ}) reads
\begin{eqnarray}
\label{L3}
&& M_1 Z(\lambda_2 , \lambda_3 , \lambda_4 | \mu_1 , \mu_2 , \mu_3) + M_2 Z(\lambda_1 , \lambda_3 , \lambda_4 | \mu_1 , \mu_2 , \mu_3)
+ M_3 Z(\lambda_1 , \lambda_2 , \lambda_4 | \mu_1 , \mu_2 , \mu_3) \nonumber \\
&&+ M_4 Z(\lambda_1 , \lambda_2 , \lambda_3 | \mu_1 , \mu_2 , \mu_3) + N_{21} Z(\lambda_0 , \lambda_3 , \lambda_4 | \mu_1 , \mu_2 , \mu_3) + N_{31} Z(\lambda_0 , \lambda_2 , \lambda_4 | \mu_1 , \mu_2 , \mu_3) \nonumber \\
&&+ N_{41} Z(\lambda_0 , \lambda_2 , \lambda_3 | \mu_1 , \mu_2 , \mu_3) + N_{32} Z(\lambda_0 , \lambda_1 , \lambda_4 | \mu_1 , \mu_2 , \mu_3) \nonumber \\
&&+ N_{42} Z(\lambda_0 , \lambda_1 , \lambda_3 | \mu_1 , \mu_2 , \mu_3) + N_{43} Z(\lambda_0 , \lambda_1 , \lambda_2 | \mu_1 , \mu_2 , \mu_3) = 0
\end{eqnarray}
and now we set $\lambda_0 = \mu_1$ and $\lambda_4 = \mu_1 - \gamma$. By doing so we find that
\begin{equation}
\left. M_1 \right|_{\lambda_0 = \mu_1, \lambda_4 = \mu_1 - \gamma} = \left. M_2 \right|_{\lambda_0 = \mu_1, \lambda_4 = \mu_1 - \gamma} = \left. M_3 \right|_{\lambda_0 = \mu_1, \lambda_4 = \mu_1 - \gamma} = 0 \; .
\end{equation}
The remaining coefficients do not vanish but from the recurrence relation (\ref{RR}) it is easy to see that
\begin{equation} 
Z(\mu_1 , \mu_1 - \gamma , \lambda | \mu_1 , \mu_2 , \mu_3) = 0 \; ,
\end{equation}
and we only need to consider the terms $M_4$, $N_{41}$, $N_{42}$ and $N_{43}$ in the Eq. (\ref{L3}). When $\lambda_0 = \mu_1$ and $\lambda_4 = \mu_1 - \gamma$
those terms simplify to 
\begin{eqnarray}
m_3 &=& \left. M_4 \right|_{\lambda_0 = \mu_1 , \lambda_4 = \mu_1 - \gamma} = c^2 a(\mu_1 - \mu_2) a(\mu_2 - \mu_1) a(\mu_1 - \mu_3) a(\mu_3 - \mu_1) \nonumber \\
\bar{m}_1 &=& \left. N_{41} \right|_{\lambda_0 = \mu_1 , \lambda_4 = \mu_1 - \gamma} = -c^2 a(\mu_2 - \mu_1) a(\mu_3 - \mu_1)
a(\lambda_1 - \mu_2) a(\lambda_1 - \mu_3) \nonumber \\
&& \qquad \qquad \qquad \qquad \;\;\; \times \frac{b(\lambda_2 - \mu_1)}{a(\lambda_2 - \mu_1)} \frac{b(\lambda_3 - \mu_1)}{a(\lambda_3 - \mu_1)}
\frac{a(\lambda_2 - \lambda_1)}{b(\lambda_2 - \lambda_1)} \frac{a(\lambda_3 - \lambda_1)}{b(\lambda_3 - \lambda_1)} \nonumber \\
\bar{m}_2 &=& \left. N_{42} \right|_{\lambda_0 = \mu_1 , \lambda_4 = \mu_1 - \gamma} = -c^2 a(\mu_2 - \mu_1) a(\mu_3 - \mu_1)
a(\lambda_2 - \mu_2) a(\lambda_2 - \mu_3) \nonumber \\
&& \qquad \qquad \qquad \qquad \;\;\; \times \frac{b(\lambda_1 - \mu_1)}{a(\lambda_1 - \mu_1)} \frac{b(\lambda_3 - \mu_1)}{a(\lambda_3 - \mu_1)}
\frac{a(\lambda_1 - \lambda_2)}{b(\lambda_1 - \lambda_2)} \frac{a(\lambda_3 - \lambda_2)}{b(\lambda_3 - \lambda_2)} \nonumber \\
\bar{m}_3 &=& \left. N_{43} \right|_{\lambda_0 = \mu_1 , \lambda_4 = \mu_1 - \gamma} = -c^2 a(\mu_2 - \mu_1) a(\mu_3 - \mu_1)
a(\lambda_3 - \mu_2) a(\lambda_3 - \mu_3) \nonumber \\ 
&& \qquad \qquad \qquad \qquad \;\;\; \times  \frac{b(\lambda_1 - \mu_1)}{a(\lambda_1 - \mu_1)} \frac{b(\lambda_2 - \mu_1)}{a(\lambda_2 - \mu_1)}
\frac{a(\lambda_1 - \lambda_3)}{b(\lambda_1 - \lambda_3)} \frac{a(\lambda_2 - \lambda_3)}{b(\lambda_2 - \lambda_3)}
\end{eqnarray}
and we are left with the equation
\begin{eqnarray}
\label{L3a}
Z(\lambda_1 , \lambda_2 , \lambda_3 | \mu_1 , \mu_2 , \mu_3) = &-& c \; a(\lambda_2 - \mu_1) a (\lambda_3 - \mu_1) a(\mu_1 - \mu_2) a(\mu_1 - \mu_3) \frac{\bar{m}_1}{m_3} Z(\lambda_2 , \lambda_3 | \mu_2 , \mu_3) \nonumber \\
&-& c \; a(\lambda_1 - \mu_1) a (\lambda_3 - \mu_1) a(\mu_1 - \mu_2) a(\mu_1 - \mu_3) \frac{\bar{m}_2}{m_3} Z(\lambda_1 , \lambda_3 | \mu_2 , \mu_3) \nonumber \\
&-& c \; a(\lambda_1 - \mu_1) a (\lambda_2 - \mu_1) a(\mu_1 - \mu_2) a(\mu_1 - \mu_3) \frac{\bar{m}_3}{m_3} Z(\lambda_1 , \lambda_2 | \mu_2 , \mu_3) \nonumber \\
\end{eqnarray}
where we have already considered the recurrence relation (\ref{RR}). Now we only need to substitute the expressions (\ref{z2}) and
(\ref{f2}) in (\ref{L3a}) in order to obtain the partition function $Z(\lambda_1 , \lambda_2 , \lambda_3 | \mu_1 , \mu_2 , \mu_3)$. With 
this procedure we automatically obtain 
\begin{equation}
\label{z3}
Z(\lambda_1 , \lambda_2 , \lambda_3 | \mu_1 , \mu_2 , \mu_3) = F_{123} + F_{132} + F_{213} + F_{231} + F_{312} + F_{321}
\end{equation}
where
\begin{equation}
\label{f3}
F_{ijk} = c^3 a(\lambda_i - \mu_2) a(\lambda_i - \mu_3) a(\lambda_j - \mu_3) b(\lambda_j - \mu_1) b(\lambda_k - \mu_1) b(\lambda_k - \mu_2) 
\frac{a(\lambda_j - \lambda_i)}{b(\lambda_j - \lambda_i)} \frac{a(\lambda_k - \lambda_i)}{b(\lambda_k - \lambda_i)} \frac{a(\lambda_k - \lambda_j)}{b(\lambda_k - \lambda_j)}
\end{equation}
The expressions (\ref{z2}) and (\ref{z3}) suggest that for general $L$ we can write the partition function as a sum over the permutation group of rather 
simple elements. In fact we can extend this analysis for general $L$ by considering $\lambda_0 = \mu_1$ and $\lambda_{L+1} = \mu_1-\gamma$
in the relation (\ref{FZ}). From the definition (\ref{mn}) we can immediately see that 
\begin{equation}
\left. M_j \right|_{\lambda_0 = \mu_1, \lambda_{L+1} = \mu_1-\gamma} = 0 \qquad \qquad j=1,\dots,L
\end{equation}
and from the recurrence relation (\ref{RR}) we obtain the property
\begin{equation}
Z(\mu_1, \mu_1-\gamma, \lambda_1 , \dots , \lambda_{L-2} | \mu_1 , \dots , \mu_L ) = 0 \;\; .
\end{equation}
In this way, for this particular choice of $\lambda_0$ and $\lambda_{L+1}$ and considering the recurrence relation (\ref{RR}), the Eq. (\ref{FZ}) reduces to
\begin{equation}
\label{LL}
Z(\lambda_1, \dots , \lambda_L) = - c \sum_{j=1}^{L} \prod_{\stackrel{k=1}{\neq j}}^{L} a(\lambda_k - \mu_1) \prod_{k=2}^{L} a(\mu_1 - \mu_k) \frac{\bar{m}_j}{m_{L}}  Z(\lambda_1, \dots , \lambda_{j-1}, \lambda_{j+1}, \dots , \lambda_L )
\end{equation}
where again we omit the dependence with the variables $\{ \mu_j \}$. In their turn the coefficients $m_L$ and $\bar{m}_j$ are given by
\begin{eqnarray}
m_{L} &=& \left. M_{L+1} \right|_{\stackrel{\lambda_0 = \mu_1}{\lambda_{L+1} = \mu_1-\gamma}} = (-1)^{L+1} c^2 \prod_{j=2}^{L} a(\mu_1 - \mu_j) a(\mu_j - \mu_1) \nonumber \\
\bar{m}_j &=& \left. N_{L+1,j} \right|_{\stackrel{\lambda_0 = \mu_1}{\lambda_{L+1} = \mu_1-\gamma}} =  
(-1)^{L} c^2 \prod_{k=2}^{L} a(\mu_k - \mu_1) a(\lambda_j - \mu_k) \prod_{\stackrel{k=1}{\neq j}}^{L} \frac{b(\lambda_k - \mu_1)}{a(\lambda_k - \mu_1)}
\frac{a(\lambda_k - \lambda_j)}{b(\lambda_k - \lambda_j)} \;\; . \nonumber \\
\end{eqnarray}
Here we remark that a similar, though not equivalent, recurrence relation has appeared previously in \cite{COL}.  
The relation (\ref{LL}) can now be iterated using the results (\ref{z2}) and (\ref{f2}), and by doing so we find
\begin{equation}
\label{zL}
Z(\lambda_1, \dots , \lambda_L) = \sum_{ \{i_1 , \dots , i_L \} \in \mathcal{S}_{L} } F_{i_1 \dots i_L}
\end{equation}
where $\mathcal{S}_L$ denotes the permutation group of order $L$ and
\begin{equation}
\label{fL}
F_{i_1 \dots i_L} = c^L \prod_{n=1}^{L} \prod_{\stackrel{j=1}{j>n}}^{L} a( \lambda_{i_n} - \mu_j ) \prod_{\stackrel{j=1}{j<n}}^{L} b( \lambda_{i_n} - \mu_j )  
\prod_{n=1}^{L-1} \prod_{m>n}^{L} \frac{a(\lambda_{i_m} - \lambda_{i_n})}{b(\lambda_{i_m} - \lambda_{i_n})} \; .
\end{equation}

Since the partition function (\ref{zL}) is given by a sum over permutations, it is not difficult to see that the invariance of $Z$
under the exchange of variables $\lambda_i \leftrightarrow \lambda_j$ is explicitly manifested in the representation 
given by (\ref{zL}) and (\ref{fL}). On the other hand, the symmetry of $Z$ under the exchange of variables $\mu_i \leftrightarrow \mu_j$
is not apparent, though the explicit evaluation of (\ref{zL}) for small values of $L$ indeed corroborates this property. 

Though it is well known that the partition function $Z$ can be written as a determinant of a $L \times L$ matrix \cite{IZE}, it
is not clear if the relations (\ref{zL}) and (\ref{fL}) can be converted into a determinant. However, since (\ref{zL}) consists
of a sum over the permutation group $\mathcal{S}_L$, it contains the same number of terms as the determinant representation \cite{IZE}.
Furthermore, the computation of the homogeneous limit $\mu_k \rightarrow \mu$ from the expressions (\ref{zL}) and (\ref{fL}) is
trivial, in constrast to what happens with Izergin-Korepin determinant representation where the evaluation of the homogeneous limit
is rather intrincated \cite{KOR1}.

\section{From functional relations to partial differential equations}
In the course of the investigation of integrable systems many connections between previously unrelated topics have emerged.
For example, the Knizhnik-Zamolodchikov (KZ) equation is a fundamental differential equation in conformal field theories and
its rich mathematical structure is manifested in the variety of topics that the KZ and its quantized version appears \cite{VAR,VAR1}.
Interesting enough the KZ equation estabilishes a connection
between two representation theories, one associated to Lie algebras and the other one associated to quantum groups \cite{VAR1,VAR2}.
Though not in the same fashion as the KZ equation connects the representation theory of Lie algebras and quantum groups,
the derivation of the functional equation (\ref{FZ}) explores a connection between the highest weight representation theories 
of the Yang-Baxter algebra and the $\mathfrak{su}(2)$ Lie algebra \cite{GAL}.
Moreover, in conformal field theory the KZ equation is a differential equation for the matrix coefficients of the product of 
intertwining operators for an affine Lie algebra $\hat{\mathfrak{g}}$, while in the case of the six vertex model with DWBC
we have a functional equation for a coefficient of the Bethe vectors.

A priori it is not clear if there exist some relation between the mentioned functional equation and the KZ equation or its quantized
version. In order to shed some light into possible connections, we aim in this section to complement the results of \cite{GAL}
by showing that the functional equation previously obtained for the partition function of the six vertex model with DWBC can be converted
into a linear partial differential equation.

Let $f$ be a complex valued function $f(z) \in \mathbb{C}[z]$ and $z = (z_1, z_2, \dots, z_n) \in \mathbb{C}^n$. For $\alpha \notin [1,n]$ we define the operator
$D_{i}^{\alpha}$ as 
\begin{equation}
\label{oper}
D_{i}^{\alpha}: \;\;\;\;\; f(z_1, \dots, z_{i}, \dots ,z_n ) \mapsto  f(z_1, \dots, z_{\alpha}, \dots ,z_n )
\end{equation}
which basically replaces the variable $z_i$ with $z_{\alpha}$.

Now we shall make use of the property (\ref{sym}) and in terms of operators $D_{i}^{\alpha}$ the functional equation (\ref{FZ}) simply reads
\begin{eqnarray}
\label{DZ}
\left\{ \sum_{1 \leq i < j \leq L} N_{ji} D_{i}^{0} D_{j}^{L+1}
+ \sum_{i=1}^{L} \left[ M_i D_{i}^{L+1} + N_{L+1,i} D_{i}^{0} \right] 
+ M_{L+1} \right\} Z(\lambda_1, \dots, \lambda_{L} ) = 0
\end{eqnarray}
which consists of a second order equation in terms of the operator $D_{i}^{\alpha}$. Moreover, as we shall demonstrate the
operator $D_{i}^{\alpha}$ possesses a differential representation when it is restricted to
the ring of polynomials. 

In order to proceed it is convenient to define the functions
\begin{eqnarray}
\bar{M}_{i} = \prod_{\stackrel{j=1}{j \neq i}}^{L+1} x_{j}^{\frac{1-L}{2}} \; M_i \;\;\;\;\;\; \mbox{and} \;\;\;\;\;\;
\bar{N}_{ji}  = \prod_{\stackrel{k=0}{k \neq i,j}}^{L+1} x_{k}^{\frac{1-L}{2}} \; N_{ji}
\end{eqnarray}
such that the Eq. (\ref{DZ}) becomes
\begin{eqnarray}
\label{DZb}
\left\{ \sum_{1 \leq i < j \leq L} \bar{N}_{ji} D_{i}^{0} D_{j}^{L+1}
+ \sum_{i=1}^{L} \left[ \bar{M}_i D_{i}^{L+1} + \bar{N}_{L+1,i} D_{i}^{0} \right] 
+ \bar{M}_{L+1} \right\} \bar{Z}(x_1, \dots, x_{L} ) = 0 \; ,
\end{eqnarray}
where the function $\bar{Z}$ defined by the Eq. (\ref{pol}) consists of a polynomial of order $L-1$ in each
variable $x_i$ separately.

In this way we can restrict the action of the operator $D_{i}^{\alpha}$ to the space of polynomials of order $m$, and this restriction is manifested
in the property
\begin{equation}
\label{cond}
\frac{\partial^k f}{\partial z_{i}^{k}} = 0  \;\;\;\;\; \mbox{for} \;\;\;\;\; k > m \;\; .
\end{equation}
Consequentely the Taylor expansion of $f$ is also truncated and convergent.

Now considering that $f = f(z_1, \dots , z_n)$ and also taking into account the Eq. (\ref{cond}) we have
\begin{eqnarray}
\label{taylor}
f &=&  f(z_1, \dots , z_{i-1} , z_{\alpha}, z_{i+1} , \dots , z_n) + \left. \frac{\partial f}{\partial z_{i}} \right|_{i = \alpha} (z_i - z_{\alpha}) \nonumber \\
&+& \frac{1}{2!} \left. \frac{\partial^2 f}{\partial z_{i}^2} \right|_{i=\alpha} (z_i - z_{\alpha})^2 + \dots
+ \frac{1}{m!} \left. \frac{\partial^m f}{\partial z_{i}^m} \right|_{i=\alpha} (z_i - z_{\alpha})^m 
\end{eqnarray}
for any $i \in [1,n]$. On the other hand, since $\alpha \notin [1,n]$ we can write 
\begin{equation}
\left. \frac{\partial^k f}{\partial z_{i}^k} \right|_{i=\alpha} =  \frac{\partial^k f(z_1, \dots , z_{i-1}, z_{\alpha}, z_{i+1}  , \dots , z_n)}{\partial z_{\alpha}^k}
\end{equation}
and the Taylor expansion (\ref{taylor}) can be rewritten as
\begin{equation}
f(z_1, \dots , z_{i-1}, z_{i}, z_{i+1} , \dots , z_n) = \left[ \sum_{k=0}^{m} \frac{(z_{i} - z_{\alpha})^{k}}{k!} \frac{\partial^{k}}{\partial z_{\alpha}^k}  \right] f(z_1, \dots , z_{i-1}, z_{\alpha}, z_{i+1} , \dots , z_n) \; .
\end{equation}
The term inside the brackets performs the operation (\ref{oper}) and we thus obtain
\begin{equation}
\label{dirrep}
D^{\alpha}_{i} = \sum_{k=0}^{m} \frac{(z_{\alpha} - z_{i})^{k}}{k!} \frac{\partial^{k}}{\partial z_{i}^k} \; .
\end{equation}

At this stage we have already gathered all the ingredients required to convert the functional equation (\ref{FZ}) into a
partial differential equation. As discussed in the section 1, the partition function (\ref{pf}) is completely determined
by the conditions (i), (ii) and (iii). Among the three conditions, the condition (iii) is the weakest one since it just determines the
leading order coefficient of the polynomial $\bar{Z}$. On the other hand the conditions (i) and (ii) play a major role in the determination 
of the partition function (\ref{pf}) and their combination is what allows us to express the Eq. (\ref{FZ}) as a linear partial differential equation.

Taking into account that the function $\bar{Z}(x_1, \dots , x_L)$ is a polynomial of degree $L-1$ in each variable $x_i$, we can substitute the representation
(\ref{dirrep}) with $m=L-1$ in the Eq. (\ref{DZb}) and we are left with
\begin{eqnarray}
\label{DP}
&& \left\{ \sum_{1 \leq i < j \leq L} \sum_{k,l=0}^{L-1} \frac{\bar{N}_{ji}}{k! l!} (x_{0} - x_{i})^{k} (x_{L+1} - x_{j})^{l}  \frac{\partial^{k+l}}{\partial x_{i}^k \partial x_{j}^l} \right. \nonumber \\
&& \left. + \sum_{i=1}^{L} \sum_{k=0}^{L-1} \frac{1}{k!} \left[ \bar{M}_i (x_{L+1} - x_{i})^{k} + \bar{N}_{L+1,i} (x_{0} - x_{i})^{k} \right] \frac{\partial^{k}}{\partial x_{i}^k} + \bar{M}_{L+1} \right\} \bar{Z}(x_1, \dots, x_{L} ) = 0 \nonumber \\
\end{eqnarray}
which consists of a linear partial differential equation of order $2(L-1)$.

\section{Concluding Remarks}
In this work we have derived a new representation for the partition function of the six vertex model
with DWBC. After Izergin's proposal of the determinant representation, this partition function has 
been rewritten in different ways over the years \cite{det1,det2,det3,det4,det5} and many connections between
this partition function and the theory of polynomials have also emerged \cite{pol,lascoux}. In particular, it was
shown in \cite{lascoux} that this partition function consists of the Schubert polynomial.

Here we have obtained a representation for this partition function in terms of a sum over the permutation
group whose possible interpretation as a determinant is not clear so far. On the other hand, the
representation (\ref{zL}, \ref{fL}) has the advantage of allowing the evaluation of the partial homogeneous limit
$\mu_k \rightarrow \mu$ in a trivial way.

Furthermore, this new representation is a direct consequence of the functional equation (\ref{FZ}) 
and Korepin's recurrence relation (\ref{RR}) removing all other requirements considered 
in \cite{GAL} and \cite{KOR} concerning the nature of the solutions. It is also worthwhile to stress
here that, though we have obtained a representation for the partition function $Z$ from the 
functional equation (\ref{FZ}), we have not solved the equation (\ref{FZ}) strictly speaking
since we have only considered the Eq. (\ref{FZ}) at the special points $\lambda_0 = \mu_1$ and
$\lambda_{L+1}= \mu_1 - \gamma$. However, a rigorous proof of the uniqueness of the solution for the
system of equations formed by (\ref{FZ}) and (\ref{RR}) seems to imply that the representation 
given by (\ref{zL}) and (\ref{fL}) indeed solves the equation (\ref{FZ}) for general values
of $\lambda_0$ and $\lambda_{L+1}$.

Concerning a second analysis of the Eq. (\ref{FZ}), we have also demonstrated in the section 4
that the requirement of polynomial solutions allows us to rewrite the functional relation (\ref{FZ})
as a linear partial differential equation. The partition function of the six vertex model with 
DWBC is known to correspond to a KP $\tau$ function \cite{foda} and also to satisfy a Toda
lattice differential equation in the homogeneous limit \cite{KOR1}. In this way we hope the linear partial differential equation
(\ref{DP}) to help shedding some light into possible connections between this partition function
and the theory of differential equations. 

\section*{Acknowledgements}
The author thanks A. Lascoux for useful discussions and correspondence.

\end{document}